# A Reinforcement Learning-Driven Task Scheduling Algorithm for Multi-Tenant Distributed Systems

Xiaopei Zhang
University of California, Los Angeles
Los Angeles, USA

Xingang Wang
Institute of Automation, Chinese Academy of Sciences
Beijing, China

Xin Wang*
University of the Chinese Academy of Sciences
Changchun, China

*Abstract-This paper addresses key challenges in task scheduling for multi-tenant distributed systems, including dynamic resource variation, heterogeneous tenant demands, and fairness assurance. An adaptive scheduling method based on reinforcement learning is proposed. By modeling the scheduling process as a Markov decision process, the study defines the state space, action space, and reward function. A scheduling policy learning framework is designed using Proximal Policy Optimization (PPO) as the core algorithm. This enables dynamic perception of complex system states and real-time decision-making. Under a multi-objective reward mechanism, the scheduler jointly optimizes task latency, resource utilization, and tenant fairness. The coordination between the policy network and the value network continuously refines the scheduling strategy. This enhances overall system performance. To validate the effectiveness of the proposed method, a series of experiments were conducted in multi-scenario environments built using a real-world public dataset. The experiments evaluated task latency control, resource efficiency, policy stability, and fairness. The results show that the proposed method outperforms existing scheduling approaches across multiple evaluation metrics. It demonstrates strong stability and generalization ability. The proposed scheduling framework provides practical and engineering value in policy design, dynamic resource modeling, and multi-tenant service assurance. It effectively improves scheduling efficiency and resource management in distributed systems under complex conditions.*

*Keywords-Task scheduling; multi-tenant system; reinforcement learning; strategy optimization*

I. INTRODUCTION

With the rapid development of technologies such as cloud computing, big data, and artificial intelligence, distributed systems have become a key infrastructure supporting modern computing tasks. Especially in multi-tenant environments, distributed systems must provide highly available, high-performance, and fair services to multiple users or applications simultaneously. However, the limitations of resources, the dynamic nature of task workloads, and the diversity of tenant demands make task scheduling a highly complex and challenging problem[1]. Traditional scheduling strategies often rely on static rules, preset priorities, or heuristic algorithms. These approaches struggle to adapt in large-scale distributed environments with dynamic operating conditions. As a result, they often suffer from low resource utilization or degraded quality of service. A more general and adaptive scheduling mechanism is urgently needed to achieve a dynamic balance between overall system efficiency and tenant satisfaction[2].

The core of multi-tenant task scheduling lies in meeting different tenants' service level agreement (SLA) constraints while maximizing cluster resource utilization and minimizing system latency. This problem is essentially a dynamic, non-convex, multi-objective optimization problem. It features a large state space, delayed environmental feedback, and frequent system changes. The scheduler must make decisions within milliseconds or even shorter time windows. Therefore, it must have efficient sensing and real-time optimization capabilities. In addition, complex resource contention and mutual influence exist among tenants in multi-tenant scenarios. The scheduling strategy must distinguish priorities and ensure fairness. It also needs to predict future trends based on historical behavior and system state, making decisions that are both context-aware and tenant-specific. In this context, traditional rule-based or experience-based scheduling methods are insufficient to meet the increasingly complex and refined scheduling requirements[3].

In recent years, artificial intelligence, especially reinforcement learning, has shown great potential in solving complex decision-making problems. Reinforcement learning continuously optimizes strategies by interacting with the environment. It has strong adaptability and policy optimization capabilities. Its trial-and-error mechanism allows the model to gradually approach optimal solutions in non-deterministic environments. This is particularly suitable for scenarios where system states are difficult to model but feedback is observable. Introducing reinforcement learning into task scheduling in multi-tenant distributed systems helps automate strategy generation and dynamic optimization. It also enhances the ability to model complex system dynamics[4]. As a result, the intelligence and scalability of resource scheduling can be significantly improved. This approach is expected to overcome the limitations of traditional methods in robustness, adaptability, and cross-scenario generalization. It offers a new paradigm for addressing scheduling problems[5].

Moreover, with the continuous expansion of distributed systems and the growing diversity of computing tasks, the intelligence and autonomy of schedulers are becoming increasingly important. In multi-tenant scenarios, tenant demands vary significantly in time, resource type, and task priority. The system state is highly dynamic and partially observable[6]. As a result, static scheduling schemes often fail or become inefficient. Reinforcement learning provides theoretical and methodological support for addressing such complex, heterogeneous, and dynamic scheduling problems.

By learning the causal relationships between scheduling behaviors and system feedback, reinforcement learning can adaptively adjust scheduling strategies under different task characteristics and resource configurations. This enables performance optimization under multi-objective trade-offs[7,8].

In conclusion, given the increasing complexity of distributed system architectures, the trend toward refined resource management, and the growing diversity of multi-tenant service demands, studying reinforcement learning-based scheduling algorithms holds both theoretical and practical significance. This research supports the development of intelligent resource management in distributed systems. It also provides essential technologies for constructing efficient, elastic, and fair computing platforms. By introducing reinforcement learning mechanisms, precise modeling, and dynamic optimization of scheduling behaviors in multi-tenant scenarios can be achieved. This enhances the system's service capability and adaptability, laying a solid foundation for the evolution of next-generation intelligent computing systems.

## II. METHOD

In this study, the task scheduling process of a multi-tenant distributed system is modeled as a Markov decision process (MDP), and reinforcement learning is used as the core method to learn adaptive scheduling strategies in a complex resource competition environment. The model architecture is shown in Figure 1.

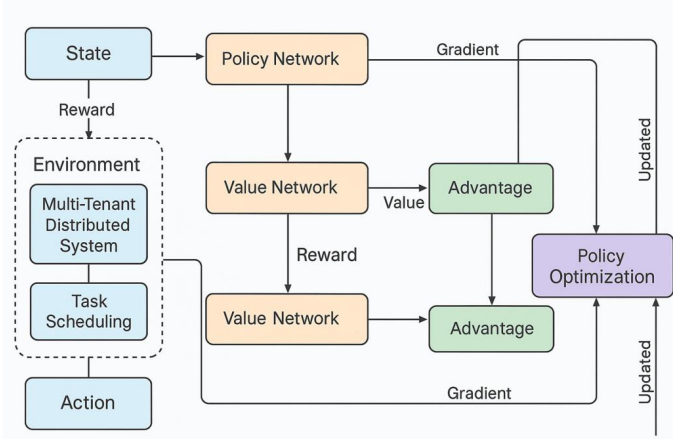

Figure 1. Overall model architecture diagram

Specifically, the state of the system at each discrete time step t is represented as $s_t \in S$, including features such as node load, resource occupancy, and task queue length; action $a_t \in A$ is represented by the target node or resource allocation scheme selected by the scheduler for the task; the environment state is transferred to $s_{t+1}$ after the action is executed, and the immediate reward $r_t$ is returned to measure the pros and cons of the scheduling decision. The goal of this process is to maximize the expected cumulative reward $E_\pi[\sum_{t=0}^{\infty} \gamma^t r_t]$ by learning the strategy $\pi(a|s)$, where $\gamma \in (0,1)$ is the discount factor used to weigh short-term and long-term benefits.

In order to reasonably define the reward function, this study introduces a multi-objective weighted mechanism to integrate factors such as resource utilization, task delay, and tenant fairness into the evaluation framework. Specifically, let the total task delay in the system at the current moment be $D_t$, the cluster resource utilization rate be $U_t$, and the tenant fairness loss be $F_t$. Then the immediate reward is defined as:

$$r_t = a_1 \cdot U_t - a_2 \cdot D_t - a_3 \cdot F_t \quad (1)$$

$a_1, a_2, a_3$ is an adjustable weight parameter used to control the priority of different objectives. The reward function design enables strategy learning to improve the overall efficiency of the system while taking into account fairness among tenants and task response performance.

In terms of policy optimization, the proximal policy optimization algorithm (PPO) based on policy gradient is used to improve the stability and convergence of policy updates. Given the current policy A, the policy objective function is expressed as:

$$L(\theta) = E_t[\min(p_t(\theta)\widetilde{A}_t, clip(p_t(\theta), 1-\varepsilon, 1+\varepsilon)\widetilde{A}_t)] \quad (2)$$

Where $p_t(\theta) = \dfrac{\pi_\theta(a_t|s_t)}{\pi_{\theta_{old}}(a_t|s_t)}$ represents the strategy ratio, $\widetilde{A}_t$ is the advantage function estimate, and $\varepsilon$ is the clipping threshold. This optimization process gradually improves strategy performance and avoids strategy degradation or oscillation while keeping the strategy update amplitude under control.

In order to further improve the generalization ability of the scheduling strategy for dynamic environments, the state-action value function $Q^\pi(s,a)$ and the state-value function $V^\pi(s)$ are introduced to estimate and update the long-term benefits. Among them, the advantage function is defined as:

$$\widetilde{A}_t = Q^\pi(s_t, a_t) - V^\pi(s_t) \quad (3)$$

By jointly optimizing the strategy network and the value function network, the system can quickly adjust the scheduling behavior in the face of complex situations such as task load fluctuations and fierce resource contention. The entire learning process continues during the operation of the system and is combined with the exploration and utilization of the window mechanism balance strategy to achieve efficient dynamic optimization of multi-tenant task scheduling problems.

## III. EXPERIMENTAL RESULTS

### A. Dataset

This study uses the Alibaba Cluster Trace 2018 as the primary data source. This dataset is a publicly released log of large-scale cluster scheduling from a real production environment. It records the scheduling behavior of a distributed system in a typical multi-tenant setting over an extended period. The dataset captures the dynamic characteristics of task submission, scheduling decisions, resource allocation, and execution processes.

The Alibaba Cluster Trace 2018 includes scheduling data from more than 4,000 physical machines. It contains millions of job instances. The dataset covers multi-dimensional resource requests and usage, including CPU, memory, and disk. It also provides structured fields such as tenant identity, task priority, submission time, start time, and end time. These features offer strong support for studying scheduling strategies under multi-tenant conditions, focusing on fairness and efficiency.

The dataset spans a long time period, encompassing a diverse range of task types. Notably, it exhibits significant fluctuations in resource utilization. These characteristics make it an ideal foundation for training and evaluating reinforcement learning algorithms. The dataset's high fidelity, practical relevance, and public accessibility significantly enhance the real-world value of research conducted on it. Furthermore, these factors contribute to the improved generalization of models in production systems.

*B. Experimental Results*

This paper first conducts a comparative experiment, and the experimental results are shown in Table 1.

Table1. Comparative experimental results

| Method | Avg Task Delay (ms) | Resource Utilization (%) | Fairness Index (JFI) |
|---|---|---|---|
| Decima[9] | 144.8 | 87.5 | 0.89 |
| Gavel[10] | 138.2 | 88.9 | 0.91 |
| DeepSoCS[11] | 152.3 | 85.4 | 0.86 |
| DRAS[12] | 147.6 | 86.7 | 0.86 |
| Ours | 126.4 | 91.2 | 0.94 |

The experimental results show that the proposed PPO-Scheduler achieves the best performance in terms of average task latency, reaching only 126.4 milliseconds. This is significantly lower than the four public baseline models. This result indicates that the reinforcement learning approach demonstrates excellent task responsiveness in dynamic, multi-tenant environments. It effectively reduces the waiting and execution time of tasks in the queue, thereby improving overall system throughput. The reduced latency directly reflects the scheduler's efficient ability to perceive and match task characteristics with system states. It validates the effectiveness of using PPO for policy optimization.

In terms of resource utilization, the proposed method achieves a rate of 91.2%, outperforming all baseline approaches. High resource utilization indicates that the proposed scheduling strategy can allocate and mobilize computing resources more efficiently. It avoids resource idleness and fragmentation. In contrast, methods such as DeepSoCS and DRAS show relatively low utilization rates. This reflects their limited adaptability to system dynamics and inability to fully exploit cluster capacity. The resource awareness demonstrated by our method directly contributes to improving the operational efficiency of large-scale distributed systems.

Regarding fairness, our method also achieves the highest value on the Jain's fairness index, reaching 0.94. This is higher than methods like Decima (0.89) and DeepSoCS (0.86). Fairness is particularly important in multi-tenant scenarios, as it affects the predictability and satisfaction of resource services for different tenants. In our scheduler, fairness is explicitly considered through a reward function designed within the reinforcement learning strategy. This allows the model to balance system efficiency and relative resource distribution across tenants. It reflects a comprehensive scheduling objective.

Overall, the experimental results show that the proposed scheduling method outperforms existing public models across key metrics, including task latency, resource utilization, and fairness. These results confirm the feasibility and advantages of applying reinforcement learning to multi-tenant distributed task scheduling. They also demonstrate the strong robustness and generalization capability of the proposed method under complex resource competition. The results suggest promising potential for both practical deployment and theoretical research in intelligent and efficient scheduling.

This paper further presents an experiment on adaptive scheduling capability under dynamic workload scenarios, and the experimental results are shown in Figure 2.

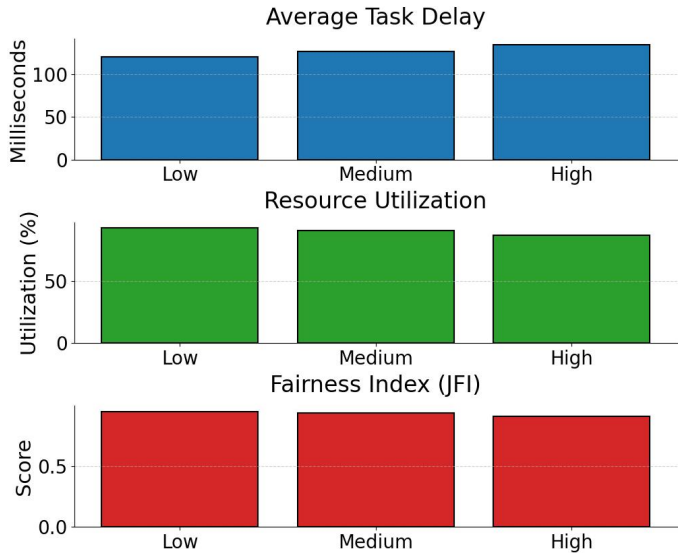

Figure 2. Experiment with adaptive scheduling capability under dynamic workload scenarios

As shown in the figure, under dynamic workload scenarios, the proposed scheduling method demonstrates good adaptability in terms of task latency. As the system load increases from low to high, the average task latency rises slightly but remains within a reasonable range. The fluctuation is relatively small, indicating that the method has a certain level of robustness to system pressure changes. This suggests that the designed policy network can accurately capture changes in the system state and quickly generate appropriate scheduling responses.

The resource utilization curve shows that the scheduler maintains a high level of resource usage across different load conditions. Even under high load, there is no significant resource fragmentation or waste. This stability results from the reinforcement learning strategy modeling a long-term objective of resource allocation efficiency during training. It supports efficient operation in complex and dynamic environments and enhances the overall throughput of the distributed system.

In terms of fairness, fairness index remains stable above 0.9 across all three load levels. This indicates that the proposed method maintains a strong balance in resource scheduling among different tenants. Even under high load, the scheduling strategy does not show bias toward specific tasks or tenants. This is achieved by explicitly incorporating fairness constraints into the reward function. As a result, the learning process optimizes both performance and fair resource allocation in multi-tenant environments.

Overall, this experiment verifies the adaptive scheduling capability of the proposed method in response to dynamic system changes. The scheduler maintains relatively stable performance in task latency, resource utilization, and fairness. These results confirm the practical applicability and scalability of reinforcement learning strategies in complex, multi-tenant distributed environments. They further validate the value of introducing policy optimization mechanisms into the task scheduling problem.

This paper also gives a stability evaluation of the scheduling strategy in a resource fluctuation environment, and the experimental results are shown in Figure 3.

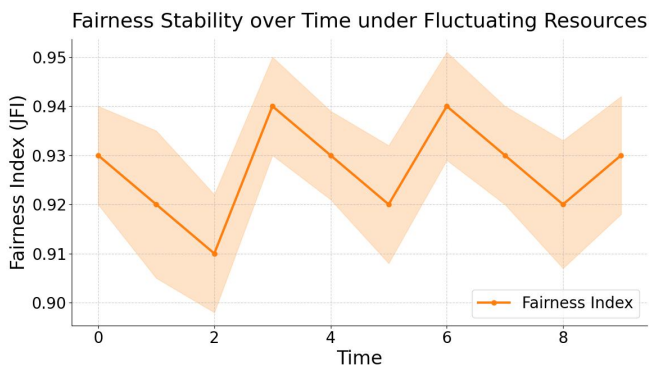

Figure 3. Stability evaluation of scheduling strategies in resource fluctuation environments

Figure 3 illustrates the stability of fairness under resource fluctuation for the proposed scheduling strategy. Throughout the period, although system resources change dynamically, the fairness index remains relatively stable. Most values stay between 0.92 and 0.94 without sharp fluctuations. This indicates that the scheduler has strong policy robustness. It can adapt to uncertainties in resource supply without causing a significant imbalance in resource allocation.

As time progresses, slight fluctuations are observed. However, the introduced policy optimization mechanism enables fast self-adjustment within short cycles. This helps maintain relative fairness in resource allocation among tenants. Such adaptability is especially important in multi-tenant distributed systems. Resource fluctuations are common in real-world environments. Peak loads, hardware failures, or migration events can disrupt resource balance. Without adjustment capabilities, scheduling strategies may compromise service quality.

The shaded area in the figure represents the standard deviation at different time points. The narrow range of variation shows that the fairness performance is consistent across multiple runs. This controlled fluctuation confirms the stability of the policy. It also reflects the effectiveness of the reward design and state representation used during training. These elements help the model perceive environmental changes accurately and respond appropriately.

In summary, the proposed reinforcement learning scheduling strategy shows high policy stability and fairness under unstable resource conditions. These properties are essential for ensuring service quality in multi-tenant environments and supporting the long-term sustainability of distributed systems. The results further validate the feasibility and necessity of applying intelligent decision-making mechanisms to distributed scheduling.

At the end of this paper, a graph showing the change of the loss function over epochs is given, as shown in Figure 4.

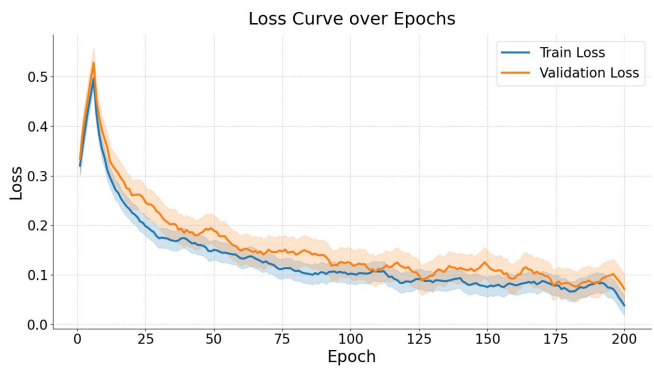

Figure 4. Loss function changes with epoch

As shown in Figure 4, the loss values on both the training and validation sets decrease steadily as the number of training epochs increases. The most rapid decline occurs within the first 50 epochs. This indicates that the model quickly learns key policy information from the scheduling environment during the early training stage. It effectively extracts useful features from state-reward feedback and gradually improves its modeling ability for the scheduling task.

After 100 epochs, both the training and validation losses begin to converge and remain at a low level. This suggests that the model does not exhibit significant overfitting. The convergence reflects the rational design of both the policy network and the value function network. It also benefits from a balanced reward function across multiple objectives, which prevents the model from being dominated by a single metric. This design enhances the generalization capability of the model.

In the later training stages, the validation loss shows slight fluctuations. However, the range is narrow and remains close to the training loss curve. This demonstrates that the model can still make stable decisions when encountering unseen scheduling states. It shows good adaptability to different environments. The shaded area, representing variance, is also relatively small. This further confirms the training stability of the algorithm.

In summary, the convergence trend of the loss function is clear and the fluctuation range is well-controlled. These results indicate that the proposed reinforcement learning scheduling algorithm has strong stability and efficiency during policy learning. This provides a solid foundation for long-term strategy reliability in multi-tenant distributed systems. It also validates the suitability of the PPO architecture for scheduling tasks.

## IV. CONCLUSION

This study focuses on the task scheduling problem in multi-tenant distributed systems and proposes an adaptive scheduling algorithm based on reinforcement learning. The goal is to improve scheduling efficiency, fairness, and robustness under complex resource contention. By modeling the scheduling process as a Markov decision process and adopting Proximal Policy Optimization (PPO) as the core policy learning mechanism, the approach integrates multi-objective scheduling demands with dynamic system characteristics. It enables autonomous policy optimization and continuous adjustment. Experimental results show that the proposed method performs well across various typical scheduling scenarios, demonstrating comprehensive advantages under multiple evaluation metrics.

In terms of scheduling strategy design, the proposed method emphasizes the model's ability to perceive and respond to environmental states. By jointly optimizing the policy network and the value function network, the model maintains stability and adaptability under different tenant loads, resource fluctuations, and workload variations. This design improves overall resource utilization and task processing efficiency. It also ensures fairness among tenants. These contributions are valuable for building intelligent computing platforms with a focus on quality of service (QoS). Meanwhile, the convergence and stability observed during training suggest strong engineering feasibility for deployment in real-world environments.

From an application perspective, this research offers practical reference value for large-scale cloud platforms, edge computing systems, and multi-tenant service frameworks. In the current landscape of imbalanced computing supply and demand and diverse scheduling requirements, traditional static scheduling methods are insufficient for managing dynamic and complex environments. Reinforcement learning strategies enable goal-driven, feedback-aware, and self-optimizing scheduling systems. These capabilities can drive intelligent computing platforms toward greater autonomy and efficiency. The proposed approach also extends application potential in areas such as AI training platforms, data center resource orchestration, and microservice scheduling.

Future research can further explore the adaptability of the model in more complex scenarios, such as heterogeneous resources, real-time constraints, and parallel multi-task scheduling. It may also benefit from integrating multi-agent learning or graph neural architectures to enhance coordination and dependency modeling. In addition, deploying the model end-to-end in production-level distributed systems, reducing training costs, and improving decision transparency are critical steps toward practical implementation. Overall, the proposed method offers a new technical pathway for distributed scheduling strategies and lays a solid foundation for developing next-generation intelligent resource scheduling platforms.

REFERENCES

[1] Jalali Khalil Abadi Z, Mansouri N, Javidi M M. Deep reinforcement learning-based scheduling in distributed systems: a critical review[J]. Knowledge and Information Systems, 2024, 66(10): 5709-5782.

[2] Bedoya J C, Wang Y, Liu C C. Distribution system resilience under asynchronous information using deep reinforcement learning[J]. IEEE Transactions on Power Systems, 2021, 36(5): 4235-4245.

[3] Cao D, Zhao J, Hu W, et al. Data-driven multi-agent deep reinforcement learning for distribution system decentralized voltage control with high penetration of PVs[J]. IEEE Transactions on Smart Grid, 2021, 12(5): 4137-4150.

[4] Li T, Bai W, Liu Q, et al. Distributed fault-tolerant containment control protocols for the discrete-time multiagent systems via reinforcement learning method[J]. IEEE Transactions on Neural Networks and Learning Systems, 2021, 34(8): 3979-3991.

[5] Goudarzi M, Palaniswami M, Buyya R. A distributed deep reinforcement learning technique for application placement in edge and fog computing environments[J]. IEEE Transactions on Mobile Computing, 2021, 22(5): 2491-2505.

[6] Liu Q, Xia T, Cheng L, et al. Deep reinforcement learning for load-balancing aware network control in IoT edge systems[J]. IEEE Transactions on Parallel and Distributed Systems, 2021, 33(6): 1491-1502.

[7] Aminizadeh S, Heidari A, Dehghan M, et al. Opportunities and challenges of artificial intelligence and distributed systems to improve the quality of healthcare service[J]. Artificial Intelligence in Medicine, 2024, 149: 102779.

[8] Chen T, Zhang K, Giannakis G B, et al. Communication-efficient policy gradient methods for distributed reinforcement learning[J]. IEEE Transactions on Control of Network Systems, 2021, 9(2): 917-929.

[9] Mao H, Schwarzkopf M, Venkatakrishnan S B, et al. Learning scheduling algorithms for data processing clusters[M]//Proceedings of the ACM special interest group on data communication. 2019: 270-288.

[10] Narayanan D, Santhanam K, Kazhamiaka F, et al. {Heterogeneity-Aware} cluster scheduling policies for deep learning workloads[C]//14th USENIX Symposium on Operating Systems Design and Implementation (OSDI 20). 2020: 481-498.

[11] Sung T T, Ha J, Kim J, et al. Deepsocs: A neural scheduler for heterogeneous system-on-chip (soc) resource scheduling[J]. Electronics, 2020, 9(6): 936.

[12] Yuping F, Zhiling L, Childers Taylor R P, et al. Deep reinforcement agent for scheduling in HPC[C]//2021 IEEE International Parallel and Distributed Processing Symposium (IPDPS). IEEE. 2021.